\documentclass[prb,twocolumn]{revtex4}%

\usepackage{graphicx}
\usepackage{amsmath}
\usepackage{ulem}
\usepackage{color}

\newcommand{\be}{\begin{equation}}
\newcommand{\ee}{\end{equation}}
\newcommand{\bea}{\begin{eqnarray*}}
\newcommand{\eea}{\end{eqnarray*}}
\newcommand{\bean}{\begin{eqnarray}}
\newcommand{\eean}{\end{eqnarray}}

\begin{document}

\draft
\title{\bf Electronic structures and transport properties of cove-edged graphene nanoribbons}

\author{David M T Kuo}

\address{Department of Electrical Engineering and Department of Physics, National Central
University, Chungli, 320 Taiwan, China}

\date{\today}

\begin{abstract}
In this comprehensive study, we undertake a thorough theoretical
examination of the electronic subband structures within cove-edged
zigzag graphene nanoribbons (CZGNRs) using the tight-binding
model. These unique nanostructures arise from the systematic
removal of carbon atoms along the zigzag edges of conventional
zigzag graphene nanoribbons (ZGNRs). Notably, CZGNRs that exhibit
intriguing band gaps can be conceptualized as interconnected
graphene quantum dots (GQDs). An essential finding of our
investigation is the inverse relationship between the size of GQDs
and the band gaps of CZGNRs, a relationship that remains
consistent regardless of the number of GQDs present. Additionally,
we delve into the examination of electron effective masses in
proximity to the edges of the first conduction subband of CZGNRs
as GQD sizes expand. We observe a significant increase in electron
effective masses as GQDs become larger, which is attributed to the
increasing similarity between larger GQDs and ZGNRs. To further
understand the practical implications, we explore the transport
properties of finite CZGNRs when connected to electrodes through
line contacts. The presence of edge defects introduces intriguing
asymmetries in the tunneling current, leading to a significant
reduction in its magnitude. Notably, we observe that the
saturation current magnitude is less influenced by the length of
CZGNRs and is instead more sensitive to the choice of materials
used for the contacted electrodes. Lastly, we investigate the
tunneling currents through GQDs featuring boron nitride textures
within the Coulomb blockade region, unveiling an irregular
staircase-like pattern in the tunneling current behavior.
\end{abstract}

\maketitle

\section{Introduction}
Since the groundbreaking discovery of two-dimensional graphene in
2004 by Novoselov and Geim [\onlinecite{Novoselovks}], a multitude
of investigations have been dedicated to the synthesis of graphene
nanoribbons (GNRs) [\onlinecite{Cai}--\onlinecite{WangX}]. The
absence of a band gap in the electronic structure of graphene
constrains its utility in optoelectronics and electronics. Prior
to 2004, theoretical examinations of armchair GNRs (AGNRs) and
zigzag GNRs (ZGNRs) demonstrated that AGNRs exhibit either
semiconducting or metallic phases, contingent on their widths,
whereas ZGNRs persist as gapless metallic phases
[\onlinecite{Nakada},\onlinecite{Wakabayashi}]. The noteworthy
predictions advanced in references
[\onlinecite{Nakada},\onlinecite{Wakabayashi}] not only propelled
the evolution of synthesis techniques for discovering novel GNR
variants but also catalyzed numerous theoretical inquiries into
these novel GNR types [\onlinecite{Cai}--\onlinecite{WangX}].

Various scenarios of GNRs involving both AGNRs and ZGNRs have been
realized using two distinct bottom-up synthesis methods:
on-surface and in-solution approaches. These methods have been
tailored to meet the demands of applications in graphene-based
electronics [\onlinecite{WangHM}]. The on-surface approach is a
powerful method for fabricating AGNRs and AGNR heterostructures.
The emergence of topological zigzag edge states in finite AGNRs
and the existence of interface-protected topological states (TSs)
in AGNR heterostructures have been experimentally and
theoretically confirmed [\onlinecite{ChenYC}, \onlinecite{Rizzo},
\onlinecite{DRizzo}]. These topological states find utility in the
realization of charge or spin quantum bits (qubits)
[\onlinecite{DRizzo}], spin current conversion devices
[\onlinecite{Kuo1}], and electron heat diodes [\onlinecite{Kuo2}].
Due to their metallic phases, not many studies have proposed GNR
devices based on ZGNRs. Nevertheless, in references
[\onlinecite{LeeYL}, \onlinecite{Florian}], authors pointed out
that cove-edge ZGNRs host interesting topological phases.

Recent advancements in the in-solution synthesis method have
demonstrated the production of graphene nanoribbons featuring
cove-shaped zigzag edges [\onlinecite{WangX}]. These cove-edged
ZGNRs (CZGNRs) display adjustable semiconducting phases with
varying band gaps [\onlinecite{WangX}]. While the topological
properties of CZGNRs have been explored using first-principle
methods [\onlinecite{LeeYL}] and tight-binding models
[\onlinecite{Florian}], a comprehensive investigation into the
electronic structures and transport characteristics of CZGNRs
across diverse scenarios, as illustrated in Fig. 1(a) and 1(b),
remains incomplete. Our intriguing discovery reveals that the band
gaps of CZGNRs depicted in Fig. 1(a) and 1(b) are not dependent on
the CZGNR lengths but are determined by the size of individual
graphene quantum dots (GQD). Notably, when edge defects are
present in CZGNRs, asymmetrical tunneling currents are observed.
Saturation currents exhibit limited sensitivity to the CZGNR
length but are significantly influenced by the coupling strengths
between the electrodes and the CZGNRs. In addition, we examine
tunneling currents through GQDs with boron nitride textures within
the Coulomb blockade region, revealing the presence of an
irregular staircase-like behavior in the tunneling currents.

\begin{figure}[h]
\centering
\includegraphics[trim=1.cm 0cm 1.cm 0cm,clip,angle=0,scale=0.3]{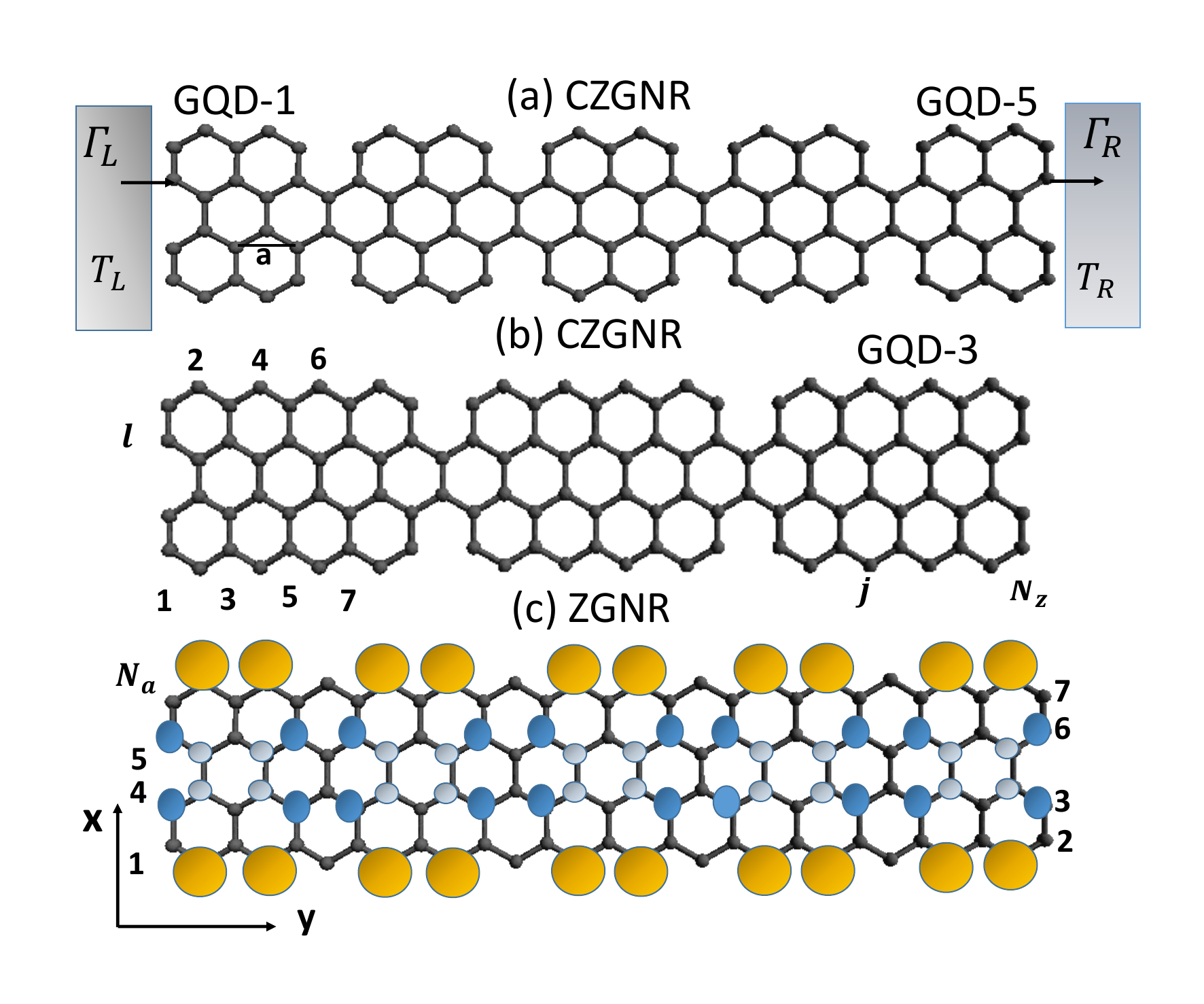}
\caption{ Schematic representation of CZGNR with $N_a = 8$ and
$N_z = 29$. (a) Armchair-edge carbon atoms of CZGNR are
interconnected with electrodes. $\Gamma_{L}$ ($\Gamma_R$)
symbolizes the electron tunneling rate between the left (right)
electrode and the leftmost (rightmost) carbon atoms at the
armchair edges. $T_{L(R)}$ represents the equilibrium temperature
of the left (right) electrode. The distances between the nearest
cove edges in (a) and (b) are $L = 3 a$ and $L = 5 a$, with the
graphene lattice constant being $a = 2.46 \AA$. Graphene quantum
dots (GQDs) in Fig. 1(a) and 1(b) can be characterized by ($N_a =
8$, $N_z = 5$) and ( $N_a = 8$, $N_z = 9$), respectively. (c)
Charge density distribution of the energy level $\varepsilon_{c} =
0.9374$ eV for ZGNR with $N_a = 8$ and $N_z = 29$. The circle's
size corresponds to the magnitude of the charge density.}
\end{figure}

\section{Calculation Methodology}
To investigate charge transport across the CZGNR coupled to
electrodes, we employ a combination of the tight-binding model and
the Green's function technique . The system Hamiltonian is
comprised of two components: $H = H_0 + H_{\text{CZGNR}}$. Here,
$H_0$ denotes the Hamiltonian of the electrodes, encompassing the
interaction between the electrodes and the CZGNR.
$H_{\text{CZGNR}}$ represents the Hamiltonian for the CZGNR and
can be expressed as follows[\onlinecite{Florian}]:

\begin{small}
\begin{eqnarray}
H_{CZGNR}&= &\sum_{\ell,j} E_{\ell,j} d^{\dagger}_{\ell,j}d_{\ell,j}\\
\nonumber&-& \sum_{\ell,j}\sum_{\ell',j'} t_{(\ell,j),(\ell', j')}
d^{\dagger}_{\ell,j} d_{\ell',j'} + h.c,
\end{eqnarray}
\end{small}
where $E_{\ell,j}$ represents the on-site energy of the $p_z$
orbital in the ${\ell}$th row and $j$th column. The operators
$d^{\dagger}_{\ell,j}$ and $d_{\ell,j}$ create and annihilate an
electron at the atom site denoted by ($\ell$,$j$).
$t_{(\ell,j),(\ell', j')}$ characterizes the electron hopping
energy from site ($\ell$,$j$) to site ($\ell'$,$j'$). The
tight-binding parameters utilized for CZGNRs are $E_{\ell,j}=0$
for the on-site energy and $t_{(\ell,j),(\ell',j')} =
t_{pp\pi}=2.7$ eV for the nearest-neighbor hopping strength.

The electron currents leaving from the electrodes are given by
\begin{eqnarray}
J&=&\frac{2e}{h}\int {d\varepsilon}~ {\cal
T}_{LR}(\varepsilon)[f_L(\varepsilon)-f_R(\varepsilon)].
\end{eqnarray}
where the Fermi distribution function of electrode $\alpha$ is
denoted as $f_{\alpha}(\varepsilon) =
1/(\exp\left(\frac{\varepsilon-\mu_{\alpha}}{k_BT}\right)+1)$. The
chemical potentials ($\mu_L = \mu + eV_{bias}$ and $\mu_R = \mu$,
with $\mu$ representing the Fermi energy of the electrodes) depend
on the applied bias. The constants $e$, $h$, $k_B$, and $T$ denote
the electron charge, Planck's constant, Boltzmann's constant, and
the equilibrium temperature of the electrodes, respectively.

In the linear response region, the electrical conductance ($G_e$)
and Seebeck coefficient ($S$) can be computed using $G_e=e^2{\cal
L}_{0}$ and $S=-{\cal L}_{1}/(eT{\cal L}_{0})$ with ${\cal L}_n$
($n=0,1$) defined as

\begin{equation}
{\cal L}_n=\frac{2}{h}\int d\varepsilon~ {\cal
T}_{LR}(\varepsilon)(\varepsilon-\mu)^n\frac{\partial
f(\varepsilon)}{\partial \mu}.
\end{equation}
Here, $f(\varepsilon)=1/(exp^{(\varepsilon-\mu)/k_BT}+1)$
represents the Fermi distribution function of electrodes. ${\cal
T}_{LR}(\varepsilon)$ signifies the transmission coefficient of a
CZGNR connected to electrodes, and it can be calculated using
formula ${\cal
T}_{LR}(\varepsilon)=4Tr[\Gamma_{L}(\varepsilon)G^{r}(\varepsilon)\Gamma_{R}(\varepsilon)G^{a}(\varepsilon)]$
[\onlinecite{Kuo3},\onlinecite{Kuo4}], where
$\Gamma_{L}(\varepsilon)$ and $\Gamma_{R}(\varepsilon)$ denote the
tunneling rate (in energy units) at the left and right leads,
respectively, and ${G}^{r}(\varepsilon)$ and
${G}^{a}(\varepsilon)$ are the retarded and advanced Green's
functions of the CZGNR, respectively. The tunneling rates are
determined by the imaginary part of the self-energy originating
from the coupling between the left (right) electrode and its
adjacent CZGNR atoms. In terms of tight-binding orbitals,
$\Gamma_{\alpha}(\varepsilon)$ and Green's functions are matrices.
For simplicity, $\Gamma_{\alpha}(\varepsilon)$ for interface
carbon atoms possesses diagonal entries with a common value of
$\Gamma_t$. Determining $\Gamma_{\alpha}(\varepsilon)$ accurately,
even using first-principle methods, is challenging
[\onlinecite{Matsuda}]. In this study, we have employed an
empirical approach to determine
it[\onlinecite{Kuo3},\onlinecite{MangnusM}].


\section{Results and discussion}
\subsection{Electronic structures of CZGNRs}
The cove-edged ZGNR (CZGNR) structures with width $N_a = 12$ can
be synthesized in reference [\onlinecite{WangX}], this study
focuses on CZGNRs with two widths characterized by $N_a = 8$ and
$N_a = 12$. It's noteworthy that CZGNRs are generated through
periodic removal of some carbon atoms from the upper and lower
zigzag edges of ZGNRs (refer to Figures 1(a) and 1(b)). To discuss
their relationships, the electron subband structures of ZGNRs with
$N_a = 8$ are presented in Figure 2(a). As for the electron
subband structures of CZGNRs with $N_a = 8$, we plot them for
varying superlattice constants $L$, defined in terms of the
graphene unit cell $a$, as depicted in Figures 2(b)-2(f). It's
important to highlight that within Figure 2(a), the states of the
first subband manifest as localized states for values of $k$
within the range $k_g = 0.738 \frac{\pi}{a} \leq k \leq
\frac{\pi}{a}$. Here, $k_g$ is computed as $k_g = 2 \times
\arccos(0.5N/(N+1))$, wherein $N = N_a/2$
[\onlinecite{Wakabayashi2}]. The corresponding eigenvalues of
$k_g$ are determined as $E_g(k_g = 0.738\pi/a) = \pm 0.529$ eV.
This finding establishes that states featuring energies within the
interval $0 \leq |E(k)| \leq 0.529$ eV correspond to localized
edge states, where wave functions exponentially decay in the
armchair direction.
\begin{figure}[h]
\centering
\includegraphics[angle=0,scale=0.3]{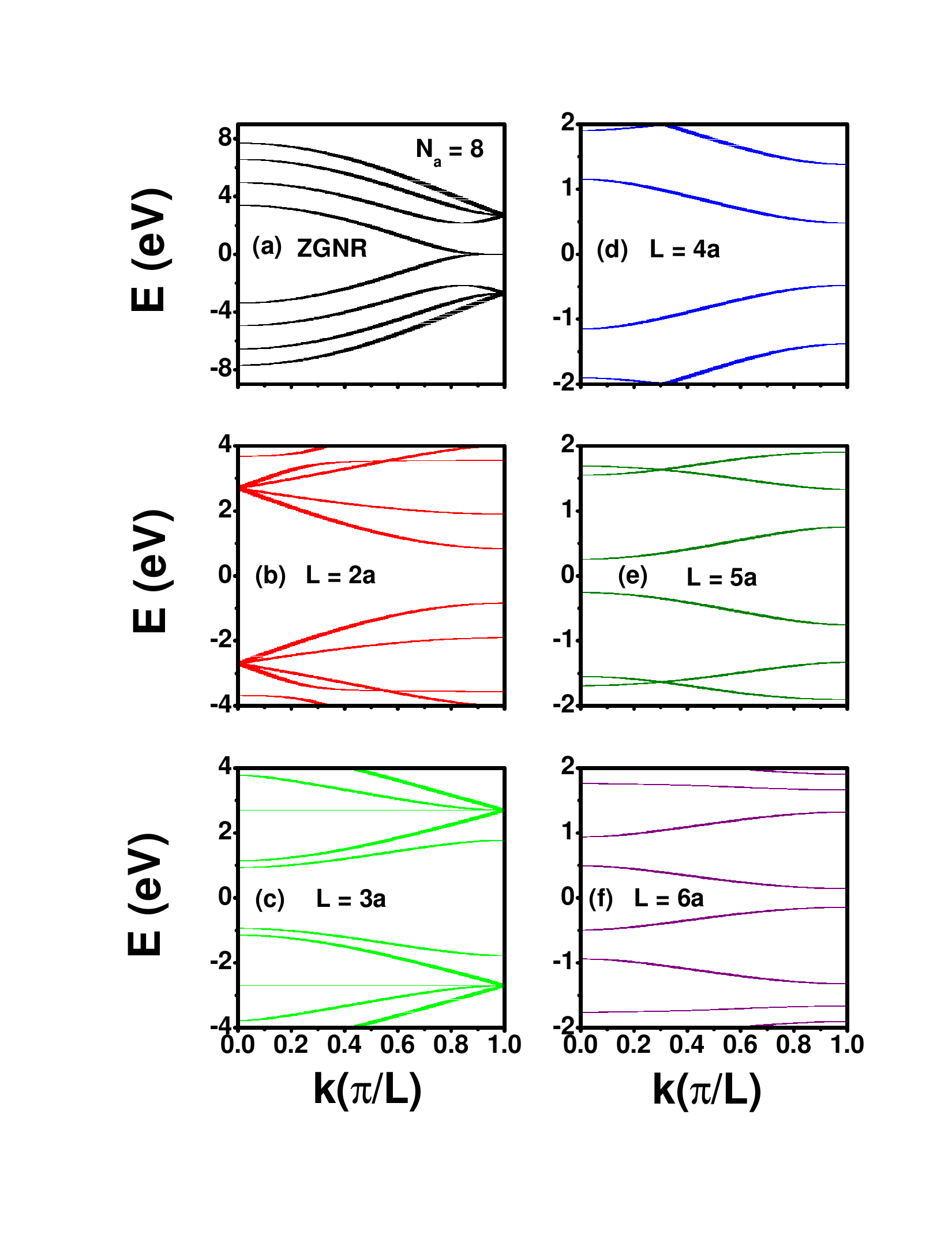}
\caption{ Electronic subband structures of CZGNRs at $N_a = 8$ for
different superlattice constants ($L$). (a) ZGNR with $N_a = 8$,
(b) $L = 2a$, (c) $L = 3a$, (d) $L = 4a$, (e) $L = 5a$, and (f) $L
= 6a$, where $a = 2.46 , \text{\AA}$.}
\end{figure}

As illustrated in Figures 2(b)-2(f), CZGNRs distinctly exhibit
semiconducting phases. Notably, the band gaps of CZGNRs diminish
with the increase of $L$. It's intriguing to observe that each GNR
with a brief zigzag segment and a width of $N_a = 8$ can be
conceptualized as an independent graphene quantum dot (GQD),which
can be characterized by $N_a = 8$ and $N_z=2L/a-1$. The increase
in $L$ effectively enlarges the GQD sizes. We found that the band
gaps of CZGNRs can be determined by the energy level separation
$\Delta = \varepsilon_{LU} - \varepsilon_{HO}$ of GQD molecules,
where $\varepsilon_{HO}$ and $\varepsilon_{LU}$ are highest
occupied molecule orbital and  lowest unoccupied molecule orbital,
respectively. Specifically, the computed gap values for CZGNRs are
$\Delta = 2.762$ eV, $\Delta = 1.875$ eV, $\Delta = 0.958$ eV,
$\Delta = 0.512$ eV, and $\Delta = 0.288$ eV, corresponding to $L
= 2a$, $L = 3a$, $L = 4a$, $L = 5a$, and $L = 6a$, respectively.
The role of GQDs plays a charge filter to diminish  the energy
levels of ZGNRs between the $\varepsilon_{HO}$ and the
$\varepsilon_{LU}$. The more detailed discussions will be given in
Fig. 4. As seen in Fig. 2(c), distinct flat subbands at $E = \pm
2.7$ eV are discernible. Conversely, for scenarios involving $L =
4a$, $L = 5a$, and $L = 6a$, minute gaps separate the first and
second conduction (valence) subbands. Meanwhile, their $\Delta <
2|E_g|=1.058$ eV, this indicates that some states near first
conduction (valence) subband lower (upper) edges could be
localized states. The outcomes depicted in Figure 2(b)-2(f)
underscore the tunability of CZGNR band gaps through adjustments
in GQD size. Particularly noteworthy is the substantial band gap
for instances where $L \leq 5a$ ($\Delta \geq 0.5$ eV), signifying
their potential utility in the realization of nanoscale
electronics and room-temperature ($k_BT = 25$~meV) thermoelectric
devices.

\begin{figure}[h]
\centering
\includegraphics[angle=0,scale=0.3]{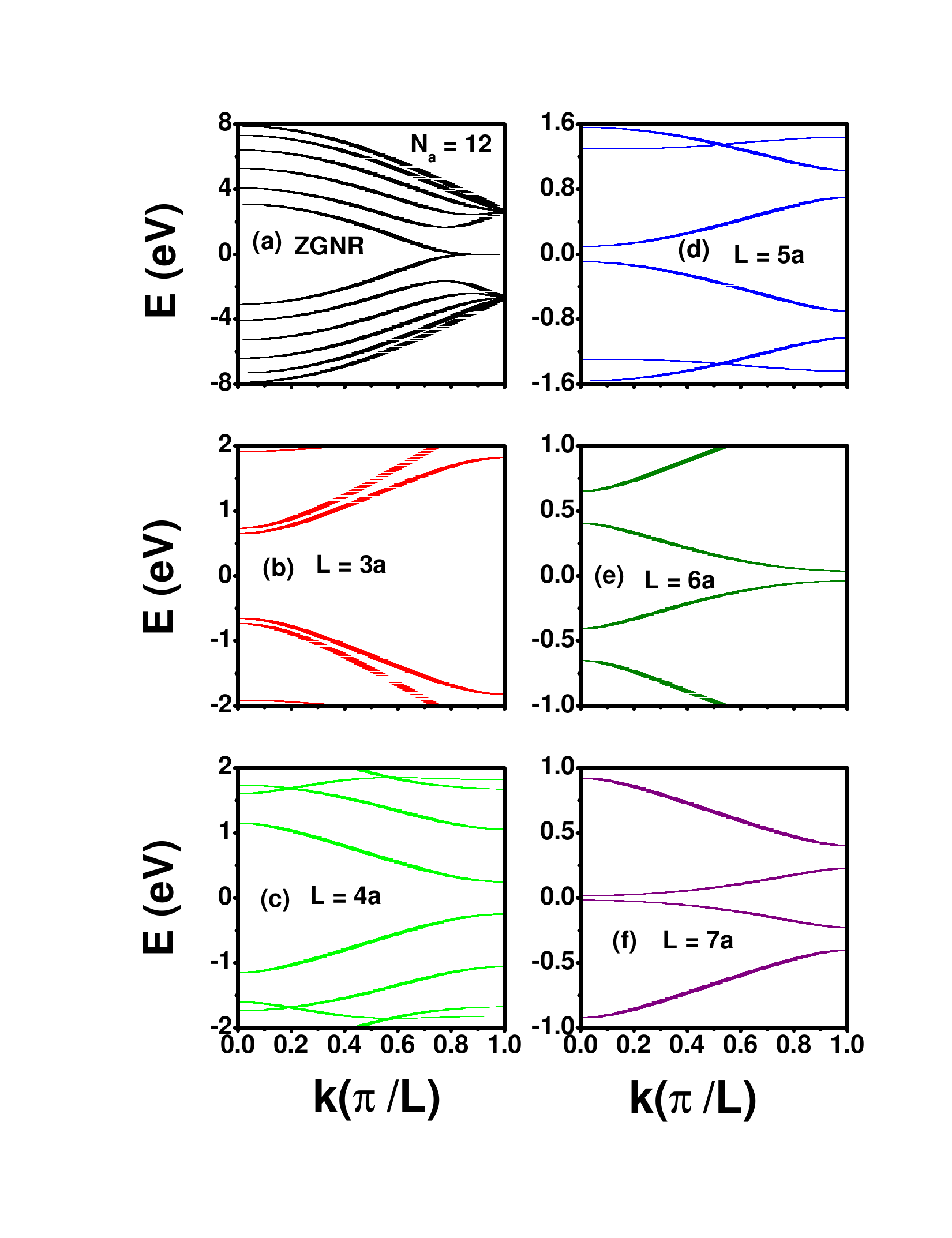}
\caption{Electronic subband structures of CZGNRs for different
superlattice constant values ($L$) at $N_a = 12$. (a) ZGNR with
$N_a = 12$, (b) $L = 3a$, (c) $L = 4a$, (d) $L = 5a$, (e) $L = 6a$
and (f) $L = 7a$, where $a=2.46 \AA$.}
\end{figure}

Subsequently, we proceed to evaluate the electronic subband
profiles of CZGNR with $N_a = 12$ across varying sizes of GQDs, as
depicted in Figure 3. Analogously to Figure 2, the subband
arrangements of ZGNRs are featured in Figure 3(a). Within the
range of $k_g = 0.717 \pi/a \leq k \leq \pi/a$, the electronic
states manifest as localized modes. Specifically, when $k_g =
0.717 \pi/a$, the corresponding energy is calculated as $E_g(k_g =
0.717\pi/a)= \pm 0.376$ eV. Much like the scenarios illustrated in
Figure 2, the band gaps exhibit a diminishing trend with the
enlargement of GQD sizes. Notably, it's observed that the band
gaps remain below $|2E_g|=0.752$ eV for $L \geq 4a$. This implies
that the states near $\varepsilon_{LU}$ and $\varepsilon_{HO}$ may
be localized states for $L \geq 4a$ when $N_a = 12$. This
situation is the same as that of $N_a = 8$. These localized states
are belonging the states of ZGNR with the flat-band structures.
The manifested characteristics are revealed in their electron
effective masses. A comprehensive overview of the band gaps and
electron effective masses for CZGNRs featuring $N_a = 8$ and $N_a
= 12$ across diverse $L$ values is presented in the subsequent
table. The computation of the electron effective mass in the
vicinity of each band gap follows the expression
$\frac{d^2E(k)}{d^2k}= \frac{75.2}{(m^{*}/m_e)(L/\AA)^2}$, where
$E$ and $k$ are dimensionless quantities. Additionally, $m^{*}$
and $L$ are expressed in terms of the electron mass ($m_e$) and
angstroms ($\AA$), respectively.

\begin{table}
\caption{Energy gap and effective mass}
\begin{tabular}{|l|l|l|l|l|l|r|}\hline \hline
{CZGNR }& \multicolumn{6}{c|}{$N_a = 8$ }\\ \hline
$L/a$&3&4&5&6&7&8\\
$\Delta/(eV)$&1.875&0.958&0.512&0.288&0.171&0.106\\
$m^{*}/m_e$ &0.35&0.235&0.237&0.265&0.282&0.323\\ \hline {CZGNR
}&\multicolumn{6}{c|}{$N_a = 12$ }\\ \hline
$L/a$&3&4&5&6&7&8\\
$\Delta/(eV)$ &1.3&0.495&0.188&0.076&0.034&0.016\\
$m^{*}/m_e$  &0.194&0.165&0.226&0.345&0.507&0.647\\ \hline \hline
\end{tabular}
\end{table}

Remarkably, the electron effective masses show an augmentation
with increasing $L$ when $L \geq 4a$, signifying a trend towards
flatter band structures as $L$ expands. It's notable that in
reference [\onlinecite{WangX}], the band gaps $\Delta$ and
electron effective mass $m^{*}$ of CZGNRs featuring $N_a = 12$
were computed for $L = 3a$ and $L = 4a$ utilizing a
first-principles approach (DFT). Specifically, they obtained
$\Delta = 1.743$ eV and $m^{*} = 0.5$ $m_e$ for $L = 3a$, and
$\Delta = 0.628$ eV and $m^{*} = 0.6$ $m_e$ for $L = 4a$. A
comparison with the results provided in the table showcases that
the band gaps computed through the tight-binding model slightly
deviate in a smaller direction compared to the DFT method. In
contrast, the electron effective masses derived from the
tight-binding method are approximately one-third of those obtained
via the DFT method. Although the values of electron effective
masses calculated by the tight binding model are not accuracy as
those of DFT, tight binding model could provide analytical
expression of electronic band structures, which can deeply reveal
the mechanism of electronic properties of
CZGNRs.[\onlinecite{Wakabayashi2}]

\subsection{Finite CZGNRs}
Due to the average size of CZGNRs being $20$ nm
[\onlinecite{WangX}], it becomes crucial to elucidate the
finite-size effects inherent in CZGNRs. In Figure 4(a), we present
computed eigenvalues of CZGNRs with $N_a = 8$ and $L = 4a$, as
functions of $N_z$ (or $N_z= (2L/a)N_{GQD}-1$), where $N_{GQD}$
signifies the GQD number ranging from 2 to 8. Notably, two
eigenvalues, namely $\varepsilon_{HO}$ and $\varepsilon_{LU}$,
exhibit a $N_{GQD}$-independent characteristic. Remarkably, a
similar intriguing characteristic is also observed at $N_a = 12,
16$. Additionally, we ascertain that the band gap depicted in
Figures 2 and 3 find their determination in $\Delta =
\varepsilon_{LO}-\varepsilon_{HO}$.

To gain deeper insights into the attributes of $\varepsilon_{HO}$
and $\varepsilon_{LU}$, we offer the eigenvalues of finite ZGNRs
with $N_a = 8$ in Figure 4(b), as functions of $N_z$, encompassing
values such as $N_z = 15, 23, 31, 39, 47, 55$, and $63$. It
emerges that four energy levels, specifically $\varepsilon_{h,2} =
-1.9024$ eV, $\varepsilon_{h,1} = -0.47905$ eV, $\varepsilon_{e,1}
= 0.47905$ eV, and $\varepsilon_{e,2} = 1.9024$ eV, within the
$|E| \le 3$ eV range, remain unaffected by variations in $N_z$.
Importantly, it is to be noted that $\varepsilon_{h,1} =
\varepsilon_{HO} = -0.47905$ eV and $\varepsilon_{e,1} =
\varepsilon_{LU} = 0.47905$ eV.

Despite the consideration of $\varepsilon_{h,1} =
\varepsilon_{HO}$ and $\varepsilon_{e,1} = \varepsilon_{LU}$ for
$L = 4a$ in Figure 4, it is pertinent to highlight that analogous
situations arise for $L = 3a$ and $L = 5a$. A numerical
verification attests that the charge density distribution of a
finite ZGNR possessing energy $\varepsilon_1$ corresponds to that
of a CZGNR with $\varepsilon_{LU}$ when their $N_a$ and $N_z$
exhibit no differences. Notably, the charge density pertaining to
$\varepsilon_{e,1}$ in a finite ZGNR with $N_a = 8$ and $N_z = 29$
is depicted in Figure 1(c). In summation, our deduction postulates
that $\varepsilon_{LU}$ (or $\varepsilon_{HO}$) embodies one of
the eigenstates of the ZGNR, featuring a specific wave function
marked by nodes at vacancy sites that correspond to the periodic
removal of carbon atoms from the zigzag edges. Upon satisfying $0
\le \varepsilon_{LU} \le |E_g(k_g)|$, it is reasonable to infer
that the states of CZGNRs in proximity to $\varepsilon_{LU}$
manifest as localized edge states with interesting magnetic
order[\onlinecite{SonYW}].

\begin{figure}[h]
\centering
\includegraphics[angle=0,scale=0.3]{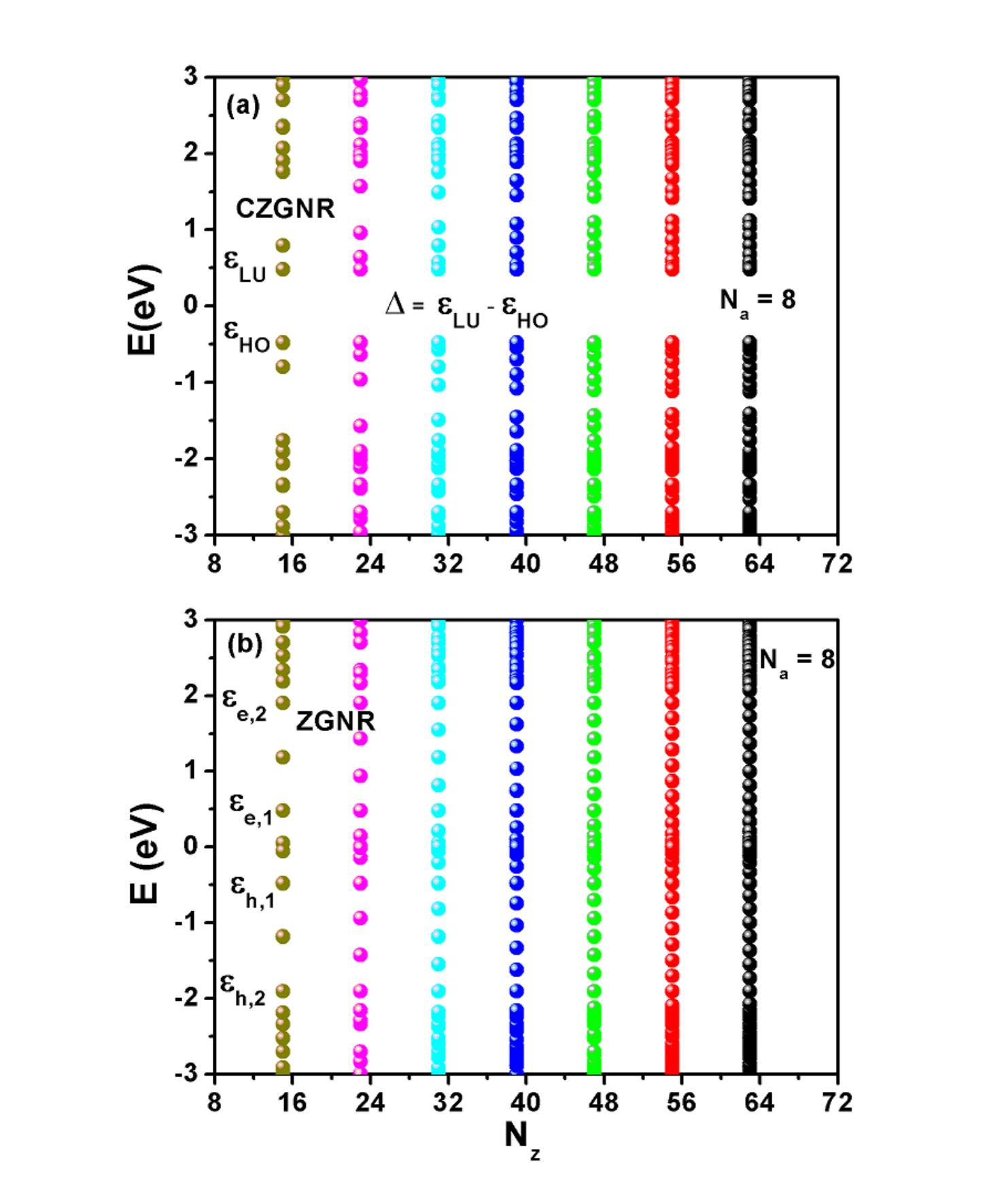}
\caption{Energy levels of finite CZGNRs and ZGNRs for various
$N_z$ values at $N_a = 8$. We have adopted $L = 4a$ corresponding
to the case of Fig. 2(d).}
\end{figure}

\subsection{Transport properties of CZGNRs in the linear response region}
When graphene interfaces with metal electrodes, the
characteristics of the contact, such as the Schottky barrier or
ohmic behavior, along with the contact geometries, wield
substantial influence over the electron transport phenomena within
graphene [\onlinecite{Matsuda}]. Illustrated in Figure 5 is the
computed transmission coefficient of CZGNRs with $N_a = 8$ and $L
= 3a$, showcasing diverse $N_z$ values at $\Gamma_t = 0.54$ eV.
This value aligns with the coupling strength of $Cu$ or $Au$
metallic electrodes [\onlinecite{Matsuda}]. For $N_z = 11, 17,
23$, the transmission coefficients reveal electron transport
marked by molecular-like traits, as each resonant energy level is
distinctly resolved. Notably, in the scenario where $N_z = 11$,
the pronounced broadening of $\varepsilon_{HO}$ and
$\varepsilon_{LU}$ due to the interaction between the electrodes
and CZGNRs is striking. This broadening phenomenon of resonant
levels diminishes as the CZGNR size increases. At instances where
two energy levels closely approach each other, they contribute to
a transmission coefficient magnitude surpassing unity.

For $N_z \ge 29$, the probabilities of transport for
$\varepsilon_{HO}$ and $\varepsilon_{LU}$ exhibit a decline with
the enlargement of $N_z$. While the spectral intensity of finite
CZGNRs proves sensitive to shifts in $N_z$, it's noteworthy that
the charge blockade region, defined by $\Delta = \varepsilon_{LU}
- \varepsilon_{HO}$, remains relatively insensitive to the
variation of $N_z$, a trend congruent with Figure 4(a). The
transmission coefficients showcased in Figure 5 furnish insight
into the electrical conductance of CZGNRs at absolute zero
temperature. Specifically, $G_e(\mu) = G_0{\cal T}_{LR}(\mu)$,
where the quantum conductance $G_0$ is expressed as $G_0 = 2e^2/h
= 1/(12.9k\Omega) = 77.5\mu S$.

\begin{figure}[h]
\centering
\includegraphics[angle=0,scale=0.3]{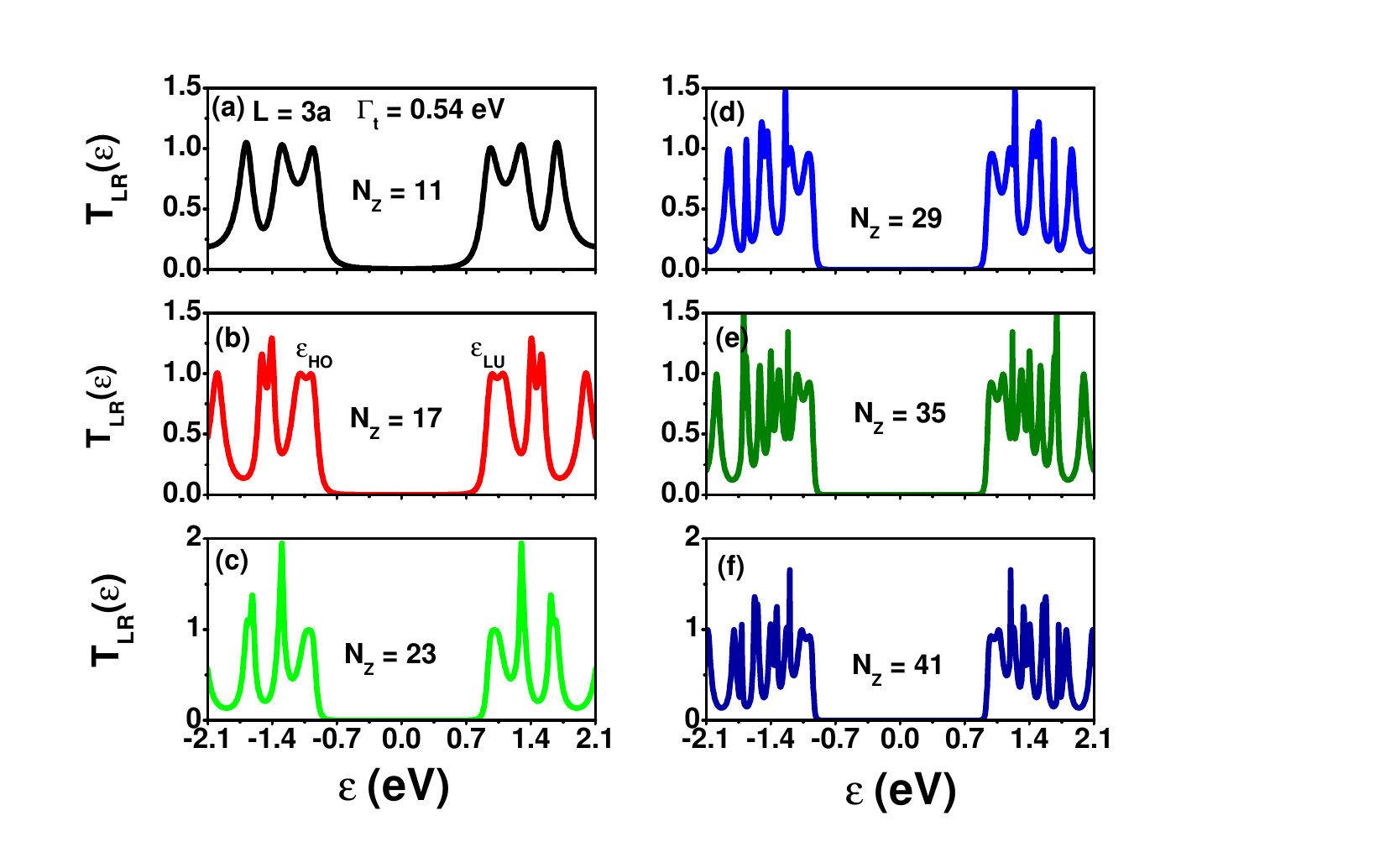}
\caption{Transmission coefficient ${\cal T}_{LR}(\varepsilon)$ of
finite CZGNRs coupled to the metallic electrodes as function of
$\varepsilon$ for various $N_z$ values at $N_a = 8$, $L = 3a$ and
$\Gamma_t= 0.54$ eV. (a) $ N_z = 11$, (b) $N_z = 17$, (c) $ N_z =
23$ , (d) $ N_z = 29$, (e) $ N_z = 35$ and (f) $N_z = 41$.}
\end{figure}

For the potential applications of nanoscale energy harvesting
[\onlinecite{KuoDMT},\onlinecite{Xu}], we present the computed
electrical conductance ($G_e$), Seebeck coefficient ($S$), and
power factor ($PF = S^2G_e$) as functions of chemical potential
values ($\mu$) for distinct $\Gamma_t$ values, all considered at
the room temperature of $T = 300$ K, as illustrated in Figure 6.
The range of $\Gamma_t$ spans from $\Gamma_t = 0.54$ eV to
$\Gamma_t = 0.81$ eV, effectively representing various metallic
electrode materials such as $Cu$, $Au$, $Pd$, and $Ti$,
respectively. In this context, the units for $G_e$, $S$, and $PF$
are pegged to $2e^2/h = 77.5\mu S$, $k_B/e = 86.25 \mu V/K$, and
$2k^2_B/h = 0.575 pW/K^2$. In the depicted graph, Figure 6(a)
specifically, it becomes evident that the electrical conductance
of subband states ($\varepsilon > \varepsilon_{LU}$) experiences
enhancement with increasing $\Gamma_t$, while the $G_e$ of subband
edge states in proximity to $\varepsilon_{LU}$ witnesses
suppression (as observed in the $G_e$ curve of Figure 6(d)).
Notably, the Seebeck coefficient ($S$) close to $\mu_{LU}$
demonstrates $\Gamma_t$-independent behavior, adhering to $S =
(\mu - \mu_{LU})/T$, wherein $\mu_{LU} = \varepsilon_{LU}$, as
portrayed in Figure 6(b). As the analysis continues, Figure 6(c)
reveals that the maximum power factor ($PF_{max}$) occurs at
$\Gamma_t = 0.54$ eV and is numerically determined as $PF_{max} =
1.1413$. It's worth noting that this maximum $PF$ value approaches
a significant milestone, reaching up to $90\%$ of the theoretical
limit, $PF_{QB} = 1.2659 \times (2k^2_B/h)$, as established for
one-dimensional (1D) systems [\onlinecite{Whitney}].

\begin{figure}[h]
\centering
\includegraphics[angle=0,scale=0.3]{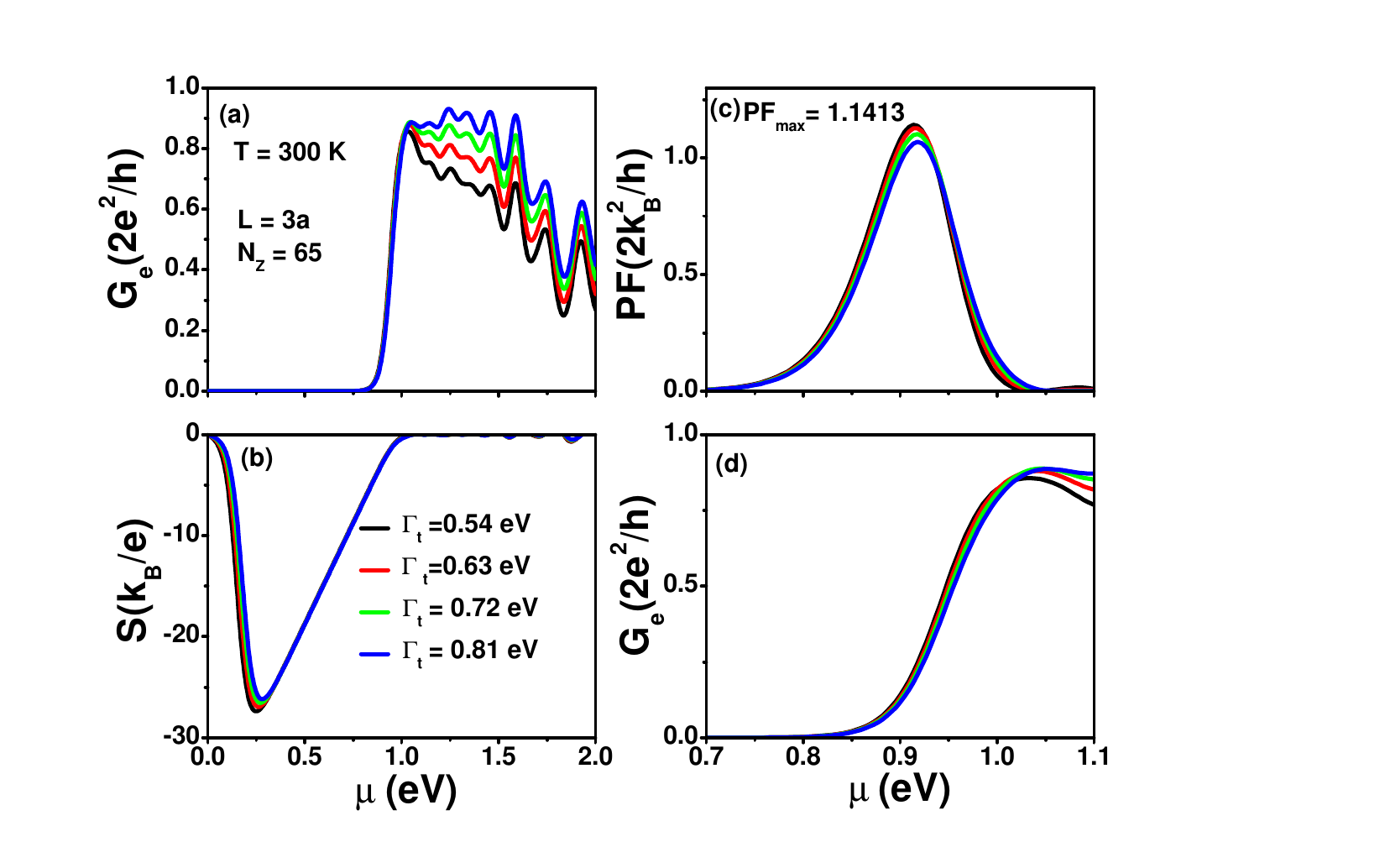}
\caption{(a) Electrical conductance $G_e$, (b) Seebeck coefficient
$S$, and (c) power factor $PF = S^2 G_e$ of CZGNR with $N_a =8$,
$N_z = 65$ and $L = 3a$ as functions of $\mu$ for various
$\Gamma_t$ values at $T = 300 K$. (d) We duplicate $G_e$ of (a) in
small chemical potential range of $0.7 eV < \mu < 1.1 eV$ to
increasing the resolution of $G_e$ with respect to $\Gamma_t$
variation. }
\end{figure}

\subsection{Transport properties of CZGNRs in the nonlinear response region}

Prior theoretical studies have demonstrated that edge defects can
significantly reduce the electron conductance of ZGNRs
[\onlinecite{Areshkin}, \onlinecite{Martins}]. In this study, we
investigate how defects at the edges influence the electron
transport properties of CZGNRs by introducing energy shifts
$\delta_{\ell,j}$ on designated defect sites. $\delta_{\ell,j}$
could be positive or negative, depending on the type of defect
[\onlinecite{Martins}]. The larger the orbital energy shift, the
stronger the effect on the electrical conductance
[\onlinecite{LiTC}]. Here, we consider the case of a negative and
large $\delta$ to investigate the effects of defects on the
electron transport of CZGNRs.

As seen in Fig. 7(a), (b), and (c), the defects occurred at edge
sites $(3, 1)$, $(1, 2)$, and $(1, 4)$ significantly influence
electron transmission coefficients when one compares these
transmission coefficients with that of the defect-free (DF)
situation. In the absence of defects, there exist 7 resonant
channels in the first conduction and valence subbands,
respectively. The asymmetrical characteristics of tunneling
currents are observed in the presence of defects as shown in Fig.
7(d), (e), and (f). In particular, the tunneling current
dramatically drops for the defect at site $(1, 4)$. This
position-dependent defect effect can be understood by their local
density of states (LDOS) of CZGNRs with defect-free. The site $(1,
4)$ has a remarkable LDOS, which is why such a defect shows a
considerable effect on the tunneling current. If defects occur in
the interior sites with small charge densities, their effects on
the tunneling current are weak (not shown here). In the case of a
defect-free scenario, we define two distinct regions: the
tunneling current cut-off region (CR), denoted as $J_{CR}$, and
the saturation region (SR), denoted as $J_{SR}$. These regions
manifest within the gap regions. Notably, the saturation current
$J_{SR}$ approximates to $2.76 \mu A$. As mentioned in the
introduction, we primarily focus on line-contacted electrodes. A
novel technique involving edge-contacted electrodes has been
developed [\onlinecite{HuangWH}]. While surface-contacted
electrodes can be readily fabricated from a device perspective,
their saturation current outputs for GNR-based devices are limited
to values smaller than one $\mu A$ [\onlinecite{LlinasJP},
\onlinecite{BraunO}].

\begin{figure}[h]
\centering
\includegraphics[angle=0,scale=0.3]{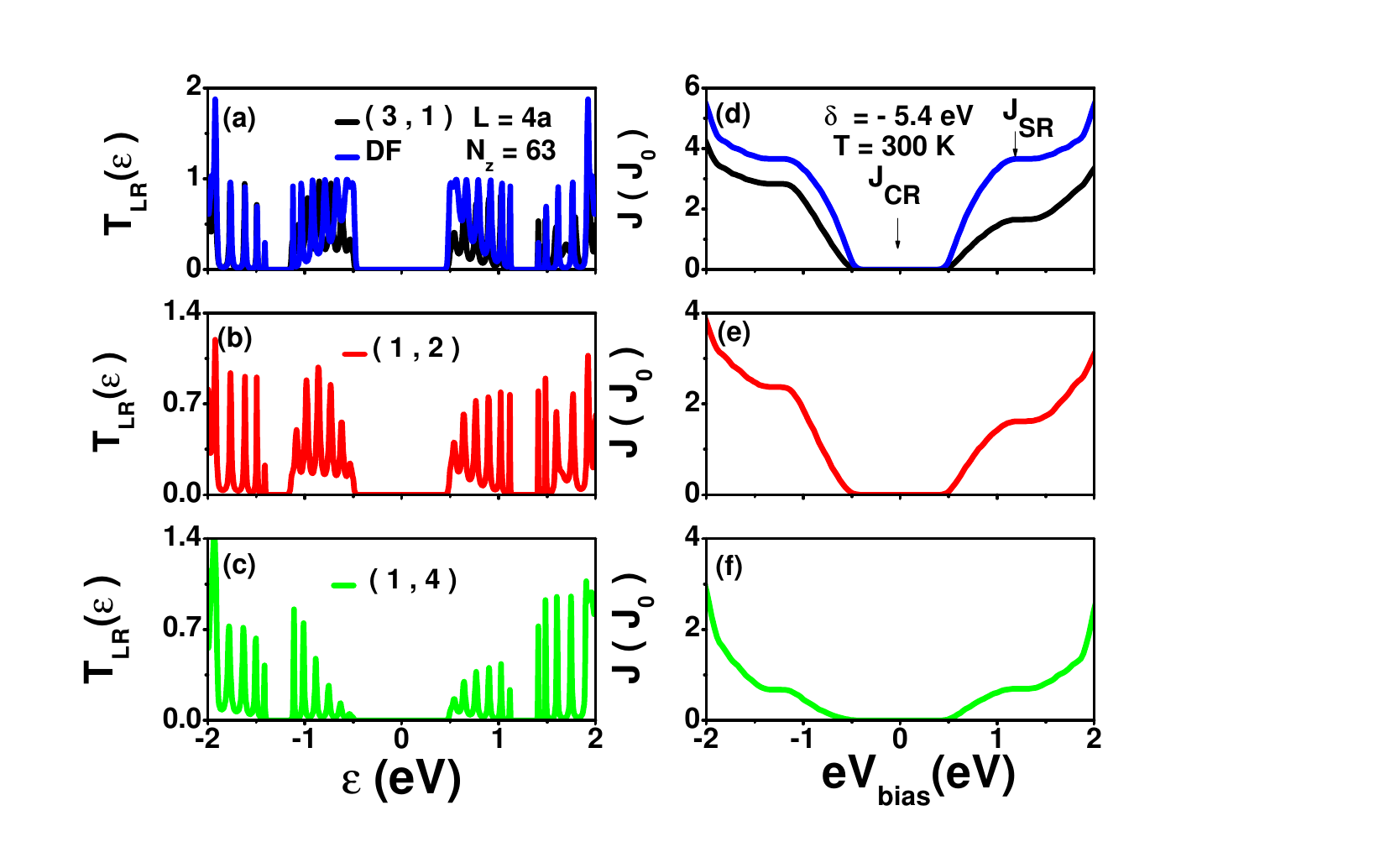}
\caption{Transmission coefficient of CZGNRs with $N_a = 8$, $N_z=
63$($L_z = 7.626$ nm) and $L = 4a $ as functions of $\varepsilon$
for different defect locations at $\Gamma_t = 0.54$ eV and $\delta
= - 5.4$ eV. (a) site (3 , 1), (b) site (1 , 2) and (c) site (1 ,
4). Tunneling current of CZGNRs as functions of applied bias
$V_{bias}$ for various defect site locations at room temperature
($T = 300$ K). (d), (e) and (f) correspond to (a), (b) and (c),
respectively. In (a) and (d), we have added an extra curve for
defect-free (DF). We have set $\mu = 0 $. Tunneling current is in
units of $J_0 = 0.773 \mu A$.}
\end{figure}

Next, we examine the impact of different $\Gamma_t$ values on the
tunneling current of CZGNRs. The transmission coefficients ${\cal
T}_{LR}(\varepsilon)$ of CZGNRs with dimensions $N_a = 8$, $N_z =
119$ (with $L_z = 14.5$ nm), and $L = 4a$ are displayed for three
distinct $\Gamma_t$ values in Figure 8(a), (b), and (c). The first
subband accommodates 15 resonant energy levels spanning a width of
$0.6647$ eV, ranging from $0.479$ eV to $1.1437$ eV. Each
resonance peak appears remarkably narrow. Notably, energy levels
further from the conduction subband edge ($\varepsilon_{LU}$)
exhibit lower transmission coefficients. The areas beneath the
${\cal T}_{LR}(\varepsilon)$ curves increase as $\Gamma_t$
increases. The patterns illustrated in Figure 8(a), (b), and (c)
underscore how electron transport through CZGNRs from
line-contacted electrodes becomes significantly modulated by the
coupling strength denoted as $\Gamma_t$, mirroring the
observations in Figure 6.

Shifting our focus to the tunneling currents depicted in Figure
8(d), (e), and (f), they correlate with the ${\cal
T}_{LR}(\varepsilon)$ curves presented in 8(a), (b), and (c),
respectively. An increase in $J_{SR}$ is noticeable with an
increase in $\Gamma_t$. The tunneling current in the saturation
region $J_{SR}$ at $\Gamma_t = 0.54$ eV is almost identical to the
blue curve in Fig. 7(d). This suggests that $J_{SR}$ is less
sensitive to variations in the length of CZGNRs when the channel
length satisfies $L_z > 7$ nm. In Figure 8(c) and 8(f), an
additional curve is included for $\Gamma_t = 2.7$ eV, which is
similar to graphene-based electrodes. The corresponding tunneling
currents exhibit a linear bias-dependent behavior (Ohmic
characteristic). Interestingly, the power outputs of these
graphene-based electrodes surpass those of metallic electrodes in
the saturation region.

\begin{figure}[h]
\centering
\includegraphics[angle=0,scale=0.3]{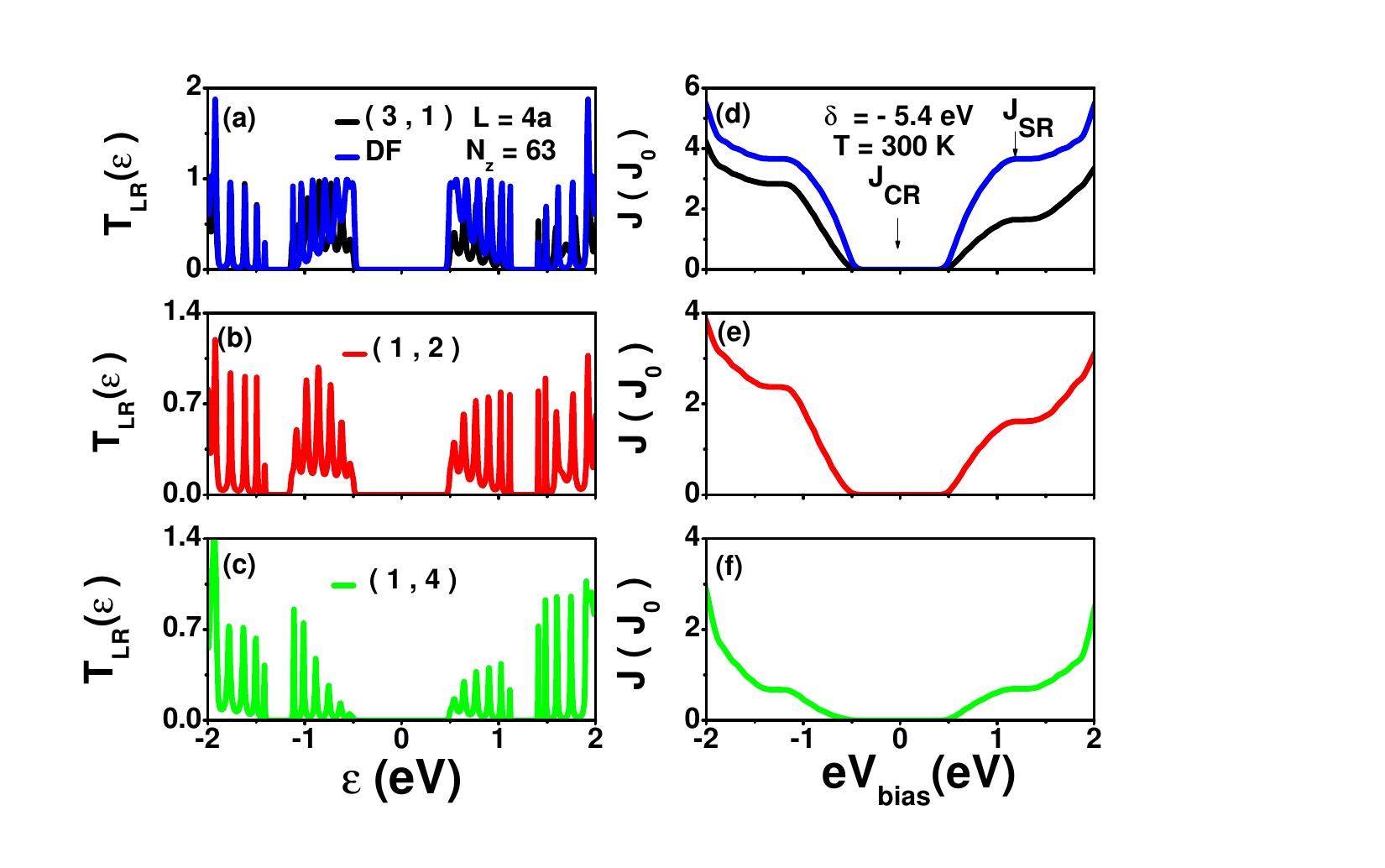}
\caption{Transmission coefficient of CZGNRs with $N_a = 8$, $N_z=
119$ ($L_z = 14.54$ nm) and $L = 4a $ as functions of
$\varepsilon$ for various $\Gamma_t$ values. (a) $\Gamma_t = 0.09$
eV, (b) $\Gamma_t = 0.27$ eV and  (c) $\Gamma_t = 0.54$ eV.
Tunneling current of CZGNRs as functions of applied bias
$V_{bias}$ for various $\Gamma_t$ values at room temperature ($T =
300$ K). (d), (e) and (f) correspond to (a), (b) and (c),
respectively. We have set $\mu = 0 $. Tunneling current is in
units of $J_0 = 0.773 \mu A$.}
\end{figure}

\subsection{Tunneling current of GQDs in the Coulomb blockade
region}

The minimization of transistors with extremely low power
consumption is always a significant concern in the semiconductor
industry[\onlinecite{WangHM},\onlinecite{ChiuKL}]. As a result,
several studies have begun to focus on electron transport through
topological states of finite armchair graphene nanoribbons and
heterostructures  in the Coulomb blockade region
[\onlinecite{Kuo1},\onlinecite{Kuo2},\onlinecite{HuangWH}]. Here,
we propose the implementation of single-electron transistors
(SETs) utilizing the schematic structure depicted in Fig. 9(a),
which differs significantly from the scenarios involving finite
AGNRs found in references
[\onlinecite{Kuo1},\onlinecite{Kuo2},\onlinecite{HuangWH}]. In
Fig. 9(a), GQDs with dimensions $N_z = 11$ and $N_a = 44$ are
connected to metallic electrodes via boron-nitride (BN) barriers.
It is worth noting that AGNRs with a width of $N_z = 11$ exhibit
very narrow energy gaps. However, when AGNRs are confined by BN
nanoribbons (BNNRs) [\onlinecite{DingY}--\onlinecite{GSSeal}],
they display semiconducting properties. To the best of our
knowledge, there has been limited investigation into the charge
transport through the upper and bottom zigzag edge segments
illustrated in Fig. 9(b). Furthermore, BNNRs serve a dual role as
both barriers and topological protectors for the localized zigzag
edge states.

\begin{figure}[h]
\centering
\includegraphics[trim=1.cm 0cm 1.cm 0cm,clip,angle=0,scale=0.3]{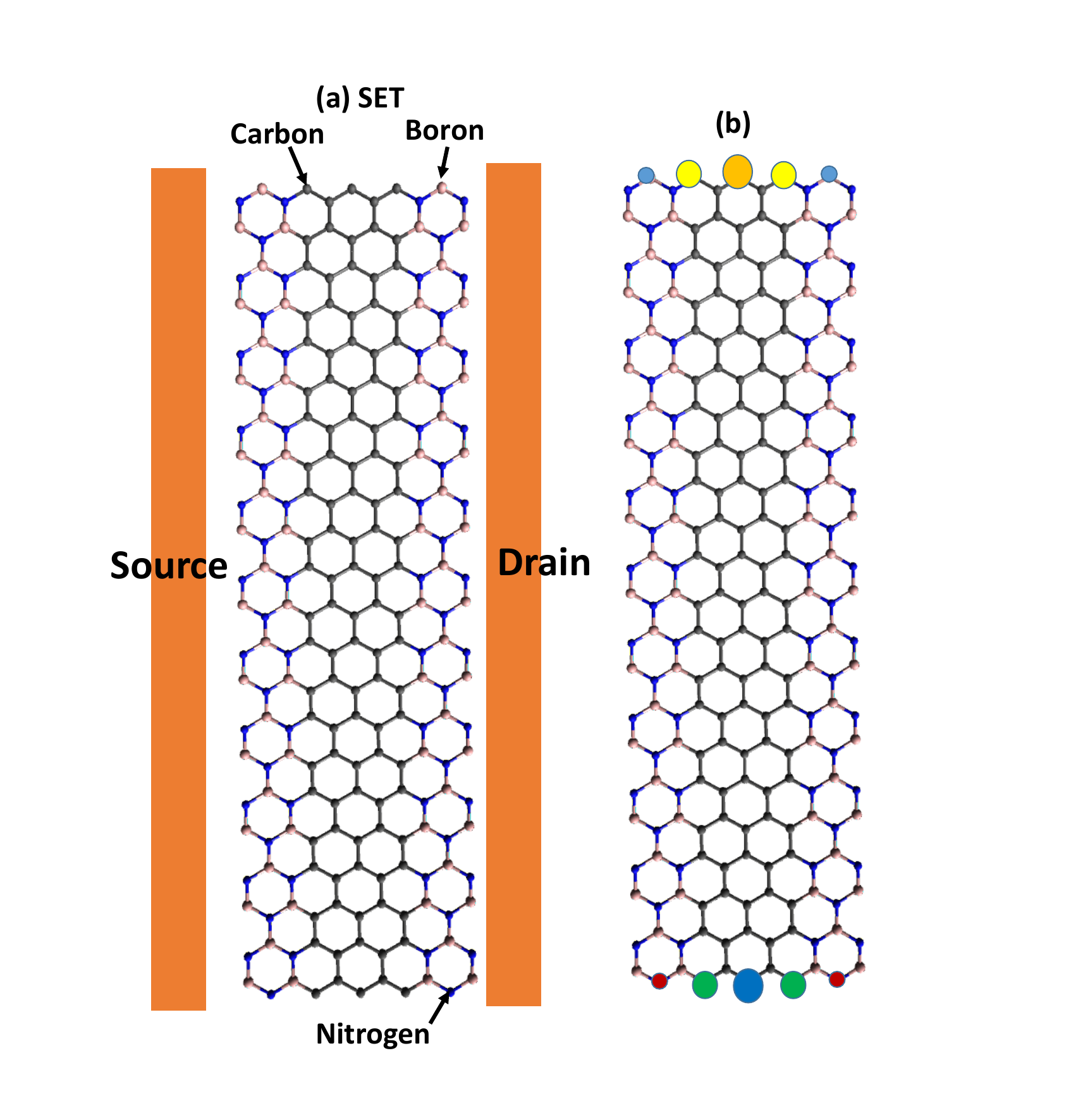}
\caption{ Schematic illustration of a single-electron transistor
(SET) utilizing a GQD with dimensions $N_a = 44$ and $N_z = 11$.
(a) Armchair-edge atoms of the GQD are interconnected with
electrodes, with boron-nitride nanoribbons serving as barriers to
enhance contact resistance. (b) Charge density distributions for
energy levels $\varepsilon_{LU} = 0.5084$ eV and $\varepsilon_{HO}
= -0.54$ eV in the GQD with $N_a = 44$ and $N_z = 11$. The charge
density is predominantly localized in the upper (bottom) zigzag
edge segment of the GQD for energy levels $\varepsilon_{LU} =
0.5084$ eV ($\varepsilon_{HO} = -0.54$ eV), with the size of the
circles representing the magnitude of the charge density.}
\end{figure}

For the sake of simplicity in our analysis, we have disregarded
variations in electron hopping strengths between different atoms
due to their relatively minor differences [\onlinecite{GSSeal}].
Specifically, we have assigned energy levels of $E_{B} = 2.329$
eV, $E_{N} = -2.499$ eV, and $E_{C} = 0$ eV to boron, nitride, and
carbon atoms, respectively. In Fig. 10, we present the computed
transmission coefficients for GQDs with BN textures at various
values of $N_a$, while keeping $N_z = 11$ and $\Gamma_t = 0.54$ eV
. Notably, a key observation is that $\Delta = \varepsilon_{LU} -
\varepsilon_{HO} = 1.048$ eV remains independent of changes in
$N_a$ (indicating a lack of size fluctuations along the armchair
edge direction). Such robustness is advantageous for the practical
implementation of SETs. The charge density distributions
corresponding to $\varepsilon_{LU} = 0.5084$ eV and
$\varepsilon_{HO} = -0.54$ eV are depicted in Fig. 9(b). As
evident from Fig. 9(b), the charge density associated with
$\varepsilon_{LU}$ ($\varepsilon_{HO}$) is predominantly confined
to the upper (bottom) zigzag segment. This observation suggests
that achieving charge transport through these zigzag edge segments
is challenging when the electrodes are connected to the zigzag
edges of GQDs with BN textures.

\begin{figure}[h]
\centering
\includegraphics[angle=0,scale=0.3]{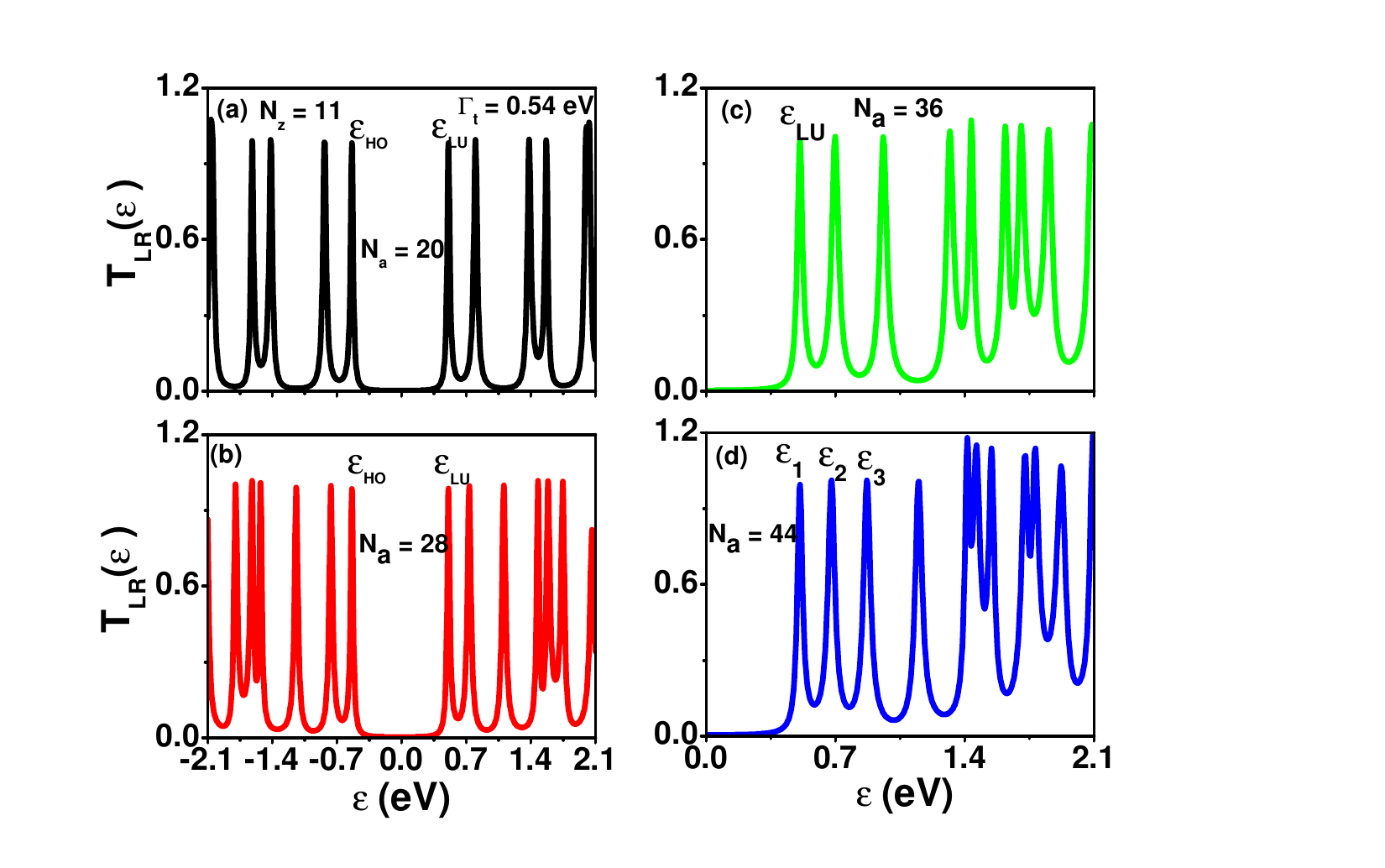}
\caption{Transmission coefficients of textured graphene quantum
dots (GQDs) with $N_z = 11$ as functions of $\varepsilon$ for
different values of $N_a$ at a fixed tunneling rate $\Gamma_t =
0.54$ eV.}
\end{figure}

To investigate charge transport within the Coulomb blockade
region, both first-principle methods like DFT
[\onlinecite{SonYW},\onlinecite{CaoT}] and the tight-binding
approach [\onlinecite{Kuo3},\onlinecite{Kuo4}] encounter technical
challenges. In our study, we have constructed an effective
Hamiltonian that accounts for multiple energy levels and electron
Coulomb interactions [\onlinecite{Kuo6}]. Specifically, we
consider an effective Hamiltonian comprising three energy levels
denoted as $\varepsilon_1 = 0.5084$ eV, $\varepsilon_2 = 0.6795$
eV, and $\varepsilon_3 = 0.8685$ eV, as shown in Fig. 10(d).
Intra-level electron Coulomb interactions are parameterized as
$U_{11} = 1.854$ eV, $U_{22} = 0.4482$ eV, and $U_{33} = 0.406$
eV, while inter-level Coulomb interactions are given by $U_{12} =
U_{21} = 0.34089$ eV, $U_{13} = U_{31} = 0.37795$ eV, and $U_{23}
= U_{32} = 0.3704$ eV. These electron Coulomb interactions are
determined by the charge densities of the energy
levels[\onlinecite{Kuo1},\onlinecite{Kuo2}].

In Fig. 11, we present the computed tunneling current within the
Coulomb blockade region. In contrast to the uniform staircase
pattern observed in the absence of interactions, we observe an
irregular staircase behavior of tunneling current due to electron
Coulomb interactions. The plateaus of these staircases result from
the interplay between energy level spacing and electron Coulomb
interactions. When $eV_{bias} > \varepsilon_3$, complex current
spectra emerge as a consequence of these interactions. The heights
of the staircases represent the magnitude of probability for each
tunneling channel. The probability associated with $\varepsilon_1$
is primarily influenced by $(1-N_{1,\sigma})$, where the
single-particle occupation number $N_{1,\sigma} = 1/3$ for small
effective tunneling rate $\Gamma_{eff}$. For $N_{1,\sigma} = 1/3$,
this explains why the tunneling current through $\varepsilon_1$ is
reduced by 1/3 within the Coulomb blockade region compared to the
non-interaction case. It's worth noting that the staircase
behavior of tunneling current is destroyed in the presence of a
large $\Gamma_{eff}$. This underscores the need for high contact
resistances in SETs.

\begin{figure}[h]
\centering
\includegraphics[angle=0,scale=0.3]{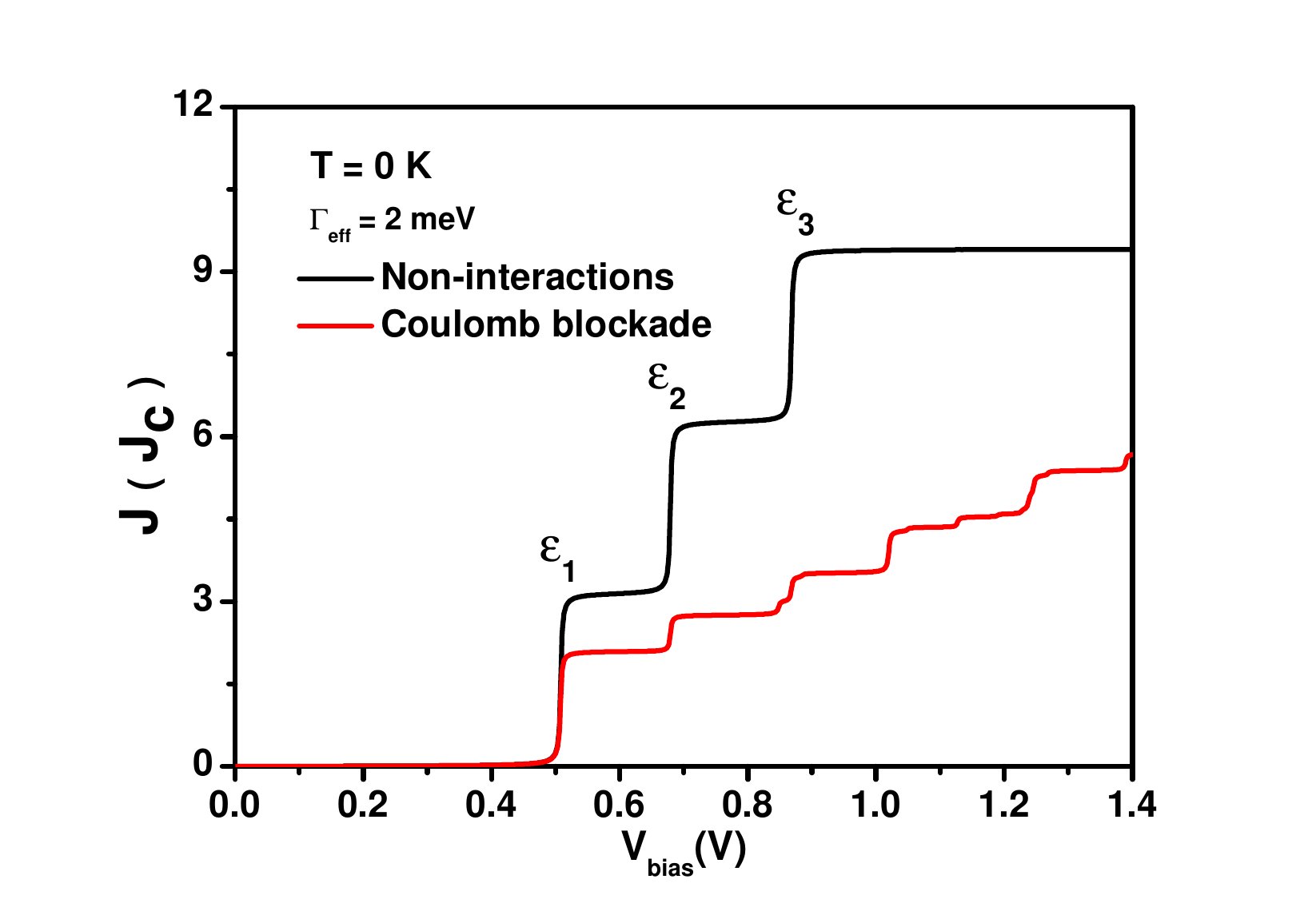}
\caption{Tunneling current of GQDs with $N_z = 11$ and $N_a = 44$
as a function of applied bias voltage ($V_{bias}$) at zero
temperature and effective tunneling rate $\Gamma_{eff,j=1,2,3} =
2$ meV (or $\Gamma_t=90$~meV). The black curve represents the
non-interaction scenario, while the red curve illustrates the
Coulomb interaction effects. Tunneling current is measured in
units of $J_c = 0.773 nA$.}
\end{figure}

\section{Conclusion}
The electronic subband structures of cove-edged zigzag graphene
nanoribbons (CZGNRs) have been systematically explored under
various conditions using the tight-binding model. Similar to
semiconducting armchair graphene nanoribbons, CZGNRs exhibit
adjustable band gaps and widths, expanding the range of promising
options in the field of graphene-based optoelectronics and
electronics applications [\onlinecite{WangHM},
\onlinecite{BraunO}].

CZGNRs can be envisioned as interconnected GQDs. The band gaps in
CZGNRs are determined by the energy level difference $\Delta =
\varepsilon_{LU} - \varepsilon_{HO}$, where $\varepsilon_{HO}$ and
$\varepsilon_{LU}$ represent the energy levels of the highest
occupied molecular orbital and the lowest unoccupied molecular
orbital of GQDs exhibiting molecular-like characteristics.
Significantly, the $\varepsilon_{HO}$ and $\varepsilon_{LU}$
values in GQDs originate from distinct eigenstates of finite
ZGNRs. The noticeable increase in the effective electron mass
within CZGNRs, as GQD sizes grow, suggests that the curvature of
the conduction subband edge increasingly resembles flat band-like
characteristics.

Moreover, we have explored the optimization of the power factor
($PF$) for CZGNRs at room temperature. Through an investigation of
CZGNRs with parameters $N_a = 8$, $N_z = 65$, and $L = 3a$, we
have achieved a remarkable $PF = 1.1413 \times \frac{2k^2_B}{h}$.
This achievement approaches $90\%$ of the theoretical limit set by
1D systems, i.e., $PF_{QB} = 1.6259 \times \frac{2k^2_B}{h}$. When
edge defects are introduced in CZGNRs, we observe intriguing
asymmetrical tunneling currents, coupled with a significant
reduction in the tunneling magnitude. The tunneling current in the
saturation region is found to be less sensitive to variations in
the length of CZGNRs ($L_z$) when $L_z > 7$ nm, but it is notably
affected by the coupling strengths between the electrodes and the
CZGNRs. Additionally, we analyze the tunneling current spectra
through GQDs with BN textures, revealing significant Coulomb
interactions in the energy level of $\varepsilon_1 =
\varepsilon_{LU}$. This indicates the potential utility of the
zigzag edge segments of GQDs with BN textures for the realization
of single electron transistors.


{}

{\bf Acknowledgments}\\
This work was supported by the National Science and Technology
Council, Taiwan under Contract No. MOST 107-2112-M-008-023MY2.

\mbox{}\\
E-mail address: mtkuo@ee.ncu.edu.tw\\

 \numberwithin{figure}{section} \numberwithin{equation}{section}

\setcounter{section}{0}
\setcounter{equation}{0} 

\mbox{}\\





\end{document}